
\input epsf
\def\slash{/\kern-6pt}
\magnification=1200
\newcount\eqnumber
\eqnumber=1
\def\neweq{\eqno(\the\eqnumber\global\advance\eqnumber by 1)}
\def\eqname#1{\xdef#1{\the\eqnumber}\neweq}
\newcount\refnumber
\refnumber=1
\def\newref{\the\refnumber\global\advance\refnumber by 1}
\def\refname#1{\xdef#1{\the\refnumber}\newref}
\pageno=1
\rightline{BNL-60062}
\rightline{Feb 1994}
\vskip.5in
\centerline{\bf Surface States and Chiral symmetry on the lattice}
\bigskip
\medskip
\centerline{Michael Creutz and Ivan Horv\'ath}
\medskip
\centerline{Physics Department}
\centerline{Brookhaven National Laboratory}
\centerline{PO Box 5000}
\centerline{Upton, NY 11973-5000}
\bigskip
\baselineskip=18pt
\centerline {ABSTRACT}
\medskip
{\narrower In a Hamiltonian formalism we study chiral symmetry
for lattice Fermions formulated in terms of Shockley surface states bound
to a wall in an extra spatial dimension.  For hadronic physics this
provides a natural scheme
for taking quark masses to zero without requiring a
precise tuning of parameters.
We illustrate the chiral anomaly as a flow of states in this extra
dimension.  We discuss two alternatives for extending the picture
to a chiral coupling of
gauge fields to such Fermions; one with a small explicit breaking of
gauge symmetry and one one with heavy mirror Fermions.
}
\vfill
This manuscript has been authored under contract number DE-AC02-76CH00016
with the U.S.~Department of Energy.  Accordingly, the U.S.~Government retains a
non-exclusive, royalty-free license to publish or reproduce the published
form of this contribution, or allow others to do so, for U.S.~Government
purposes.
\eject

\noindent{\bf I. Introduction}

Chiral symmetry has long played an essential role in particle theory.
The pion is made of the same quarks as the rho meson, yet its mass
is considerably less.  The canonical explanation says that if quarks were
massless, then the pions would be Goldstone bosons arising from
the spontaneous breaking of an underlying chiral symmetry.  Indeed,
pure gauge interactions are helicity conserving, and thus both the number of
left and right handed massless quarks are separately
conserved.  Through confinement
into the physical states of baryons and mesons, this symmetry is
spontaneously broken.  A host of predictions from current algebra
are based on this picture [\refname{\currentalgebraref}].

These issues are complicated by the presence
of ``anomalies.''  Ultraviolet divergences make it impossible,
even in perturbation theory, to
conserve simultaneously all axial vector currents associated with chiral
symmetry, and the vector current of electric charge.  As current
conservation is crucial to our understanding of gauge symmetries,
we must conserve the vector current, implying that
the axial symmetry cannot be exact.

One consequence of the anomaly is that there is one less Goldstone boson than
naive counting would suggest.  For two flavors of light quarks, of the
four ways to form pseudoscalar mesons there
are three light pions while the eta remains heavier.
With $SU(3)$ flavor symmetry, it is
the $\eta^\prime$ which is anomalously heavy compared to the other
pseudoscalars.

Over the last 20 years, lattice techniques have become the dominant
method for the study of non-perturbative quantum field theory.  The
understanding
of chiral symmetry in this framework has remained an infamous and
elusive goal for the entire
period.  For a review
see Ref.~[\refname{\petcherref}].
The problems with lattice chiral symmetry
are intricately entwined with the chiral anomaly.  Indeed,
as the lattice is a regulator, anomalies can only arise through
non-chirally symmetric terms in the underlying action or Hamiltonian.
Simple prescriptions without such terms are plagued with the so
called ``doubling'' problem, wherein extra, usually unwanted,
Fermionic species appear to cancel any anomalies in the theory.

Chiral issues arise in an even more fundamental way with the weak
interactions.  Here parity
violation seems to maximally differentiate
between left and right handed Fermions.  While lattice methods have
been dominantly applied to the strong interactions, there are reasons
to desire a lattice formulation of the weak interactions
as well.  In particular, the lattice is the best founded non-perturbative
regulator, and thus provides an elegant framework for the definition
of a quantum field theory.  Even though the smallness of the electromagnetic
coupling makes non-perturbative effects quite small in
the electroweak theory, at least in principle we would like
a rigorous formulation.  While exceptionally
small, some interesting non-perturbative
phenomena are directly
related to the anomaly, such as the prediction that baryons can
decay through the instanton mechanism of 't Hooft [\refname{\thooftref}].

The last year has seen considerable theoretical activity on the use of
surface states as a basis for a theory of chiral lattice
Fermions [\refname\kaplanref].
Here one envisions
our four dimensional world as an interface  embedded in a five dimensional
underlying space.  With appropriate conditions on the model, low energy
Fermionic states are bound to this wall.  With these being the
only low energy states, we have an effective chiral theory
on the interface.  The prime purpose of this paper is to investigate
these models in a Hamiltonian language and attempt to obtain a physical
understanding of how the anomaly works.  Jansen [\refname\jansenref] has
performed a similar study, and this should be considered as an extension
of that work.  A preliminary discussion of some
of our ideas is contained in Ref.~[\refname\latref].
\bigskip

\bigskip
\noindent {\bf II. Massless Fermions and the anomaly}

Chiral symmetry is intimately tied with Lorentz invariance.  A massive particle
of spin $s$ has $2s+1$ distinct spin states which mix under a general
Lorentz transformation.  The helicity of a massless particle,
on the other hand, is frame invariant.  Indeed, for free
particles one can write down local fields which create or destroy
just a single helicity state.  For spin 1/2 Fermions coupled minimally
to gauge fields, their helicity remains naively conserved.

In one space dimension the roles of left and right handed helicities
are replaced by left and right moving particles.  Since an observer
cannot go faster than light, he can never overtake a massless particle
and a right mover will be so in all Lorentz frames.

The fact that Lorentz invariance is crucial here provides another warning
that chiral symmetry on the lattice will be difficult.  Indeed, lattice
formulations inherently violate the usual space time symmetries.
Chiral issues should only be expected to be useful for states of low
energy which do not see the underlying lattice structure.

While separate phase rotations of left and right handed massless Fermions
give a formal symmetry of a continuum gauge theory, well known
anomalies arise through the divergences of quantum field theory.
In particular, there is the famous triangle diagram where a virtual
Fermion loop couples an axial
vector current to two vector currents.  This diagram cannot
be regulated so that both the vector and axial vector currents are conserved.
In two space time dimensions the analogous problem arises with a simple
bubble diagram connecting a vector and an axial vector current.

The fact that the anomaly must exist can be intuitively argued in
analogy with band theory in solid state physics.  With massive Fermions
the vacuum has a Fermi level midway
in a gap between the filled Dirac sea and a continuum of positive energy
particle states.  This represents an insulator.
As the mass is taken to zero, the gap closes and
the vacuum becomes a conductor.  External gauge fields applied
to this conductor can induce currents.
For a specific example, consider
a one space dimensional world compactified into a ring.  A changing
magnetic field through this ring will induce currents, changing the relative
number of left and right moving particles.
Without the anomaly, transformers would not work.

In this problem, physics should be periodic in the amount of flux
through the ring.  This is a one dimensional analog of the
periodicity of four dimensional non-abelian gauge theories as one
passes through topologically non-trivial configurations [\refname\cmtref].
The latter case with the standard model gives rise to a non-conservation
of the baryon current [\thooftref].

With the one dimensional ring, the strength of the field characterizes
the phase that a charged particle acquires in running around the ring.  As
this net phase adiabatically increases, the individual Fermionic energy
levels shift monotonically.  As one adds another unit of flux
through the ring, one filled right moving level from the Dirac sea moves, say,
up to positive energy, while one empty left moving level drops into the
sea, leaving a hole.  This induces a net current carried by a
right moving particle and a left moving antiparticle.
This way of visualizing how the anomaly works was nicely discussed
some time ago [\refname\ambjornref].  We
will find a similar picture
when we investigate the lattice models.

{\noindent \bf III. The doubling problem}

The essence of the lattice doubling problem already appears with
the simplest Fermion Hamiltonian in one space dimension
$$
H=iK\sum_j a_{j+1}^\dagger a_j - a_j^\dagger a_{j+1}. \eqname{\hoppereq}
$$
Here $j$ is an integer labeling the sites of an infinite
chain and the $a_j$ are Fermion annihilation operators satisfying
standard anticommutation relations
$$
\left[ a_j, a_k^\dagger\right]_+\equiv a_j a_k^\dagger+ a_k^\dagger a_j
=\delta_{j,k}. \neweq
$$
The bare vacuum $\vert 0 \rangle$ satisfies $a_j \vert 0 \rangle=0.$  This
vacuum is not the physical one, which contains a filled Dirac sea.
We refer to $K$ as the hopping parameter.
Energy eigenstates in the single Fermion sector
$$
\vert \chi\rangle=\sum_j \chi_j a_j^\dagger \vert 0 \rangle \neweq
$$
can be easily found in momentum space
$$
\chi_j= e^{iqj} \chi_0. \neweq
$$
where $0 \le q < 2\pi$.  The result is
$$
E(q)=2K \sin(q). \neweq
$$
The physical vacuum has the negative energy states filled
to form a Dirac sea.  This is sketched in Fig.~(1).

\midinsert
\epsfxsize=.7\hsize
\noindent \narrower {{\bf Fig.~(1)}
The spectrum of Fermions hopping along a line
with the Hamiltonian in eq.(\hoppereq).}
\endinsert

If we consider a Fermionic wave packet produced from small momentum $q$,
then since the group velocity $dE/dq$ is positive in this region, it will
move to the right.  On the other hand, a wave packet produced from
momenta in the vicinity of $q\sim \pi$ will be left moving.  The essence
of the Nielsen Ninomiya theorem [\refname{\nogoref}] is that we
must have both types of excitation.  The
periodicity in $q$ requires the dispersion relation to have
an equal number of zeros with positive and negative slopes.

The recent attempts to circumvent this result add to the spectrum
an infinite number of
additional states at high energy[\refname\nnref].
The idea is to have a mode with
$E=2K\sin(q)$ still exist at small $q$, but then become absorbed
in an infinite band of states before $q$ reaches $\pi$.  If the
band is truly infinite, then the extra state does not have to reappear
as the momentum increases to $2\pi$.
In the domain wall picture, this infinite tower
of states is represented by a flow into the extra dimension.

\bigskip
\noindent{\bf IV. The Wilson approach}

In this section we review Wilson's scheme for adding a non-chirally
symmetric term to remove the doublers appearing in a naive lattice
transcription of the Dirac equation.  We do this in some detail
because the general behavior of the Wilson-Fermion Hamiltonian
will be central to our later construction of surface modes.
To keep the discussion
simple, we work in one dimension with a two component
spinor
$$
\psi=\pmatrix{a\cr b\cr}.
\neweq
$$
  The most naive lattice Hamiltonian
begins with the simple hopping case of Eq.~(\hoppereq) and adds in
the lower components and a mass term to mix the upper and lower components
$$\eqalign{
H=i&K\sum_j
a_{j+1}^\dagger a_j - a_j^\dagger a_{j+1}
-b_{j+1}^\dagger b_j + b_j^\dagger b_{j+1}\cr
 +&M\sum_j a^\dagger_j b_j + b^\dagger_j a_j.\cr
} \neweq
$$
Introducing Dirac matrices
$$
\gamma_0=\pmatrix{0&1\cr
                   1&0\cr},
\ \ \gamma_1=\pmatrix{0&-1\cr
                             1&0\cr}
\neweq
$$
and defining
\def \psibar{\overline\psi}

$\psibar=\psi^\dagger\gamma_0$,
we write the Hamiltonian more compactly as
$$
H=\sum_j iK(\psibar_{j+1} \gamma_1 \psi_j-\psibar_j \gamma_1 \psi_{j+1})
+M\sum_j\psibar_j \psi_j.
\neweq
$$

As before, the single particle states are easily
found by Fourier transformation and satisfy
$$
E^2=4K^2 \sin^2(q)+M^2 \neweq
$$
This spectrum is sketched in Fig.~(2).
Again, we are to fill the negative energy sea.

\midinsert
\epsfxsize=.7\hsize
\noindent \narrower {{ \bf Fig.~(2)} The spectrum of naive Fermions in
 one dimension.}
\endinsert

Naive chiral symmetry is implemented through distinct phase
rotations for the upper and lower components of $\psi$.  The mass term
mixes these components and opens up a gap in the spectrum.  The doublers
at $q\sim\pi$, however, are still with us.

To remove the degenerate doublers, we make the mixing of the
upper and lower components momentum dependent.  A simple way of doing this
was proposed by Wilson [\refname\wilsonref].  In our language, we add
one more term to the Hamiltonian
$$\eqalign{
H=i&K\sum_j
a_{j+1}^\dagger a_j - a_j^\dagger a_{j+1}
-b_{j+1}^\dagger b_j + b_j^\dagger b_{j+1}\cr
 +&M\sum_j a^\dagger_j b_j + b^\dagger_j a_j\cr
-r&K \sum_j a_j^\dagger b_{j+1}+b_j^\dagger a_{j+1}
 +b_{j+1}^\dagger a_j+a_{j+1}^\dagger b_j \cr
=&\sum_j K(\psibar_{j+1} (i\gamma_1-r) \psi_j-\psibar_j (i\gamma_1+r)
 \psi_{j+1})+\sum_jM\psibar_j \psi_j.\cr
}
\eqname{\Hamiltonianeq}
$$
Now the spectrum satisfies
$$
E^2=4K^2 \sin^2(q)+(M-2rK\cos(q))^2. \neweq
$$
This is sketched in Fig.~(3).

Note how the doublers at $q\sim \pi$
are increased in energy relative to the states at $q\sim 0$.  The physical
particle mass is now $ m=M-2rK $ while the doubler is at $M+2rK$.

\midinsert
\epsfxsize=.7\hsize
\noindent \narrower {{ \bf Fig.~(3)} The spectrum of Wilson Fermions
   in one dimension.}
\endinsert

When $r$ becomes large, the dip in the spectrum of
Fig.~(3) near $q=\pi$ actually becomes
a maximum.  This is irrelevant for our discussion, although we note that
the case $r=1$ is somewhat special.  For this value, the matrices
$i\gamma_1\pm r$, which
determine how the Fermions hop along the lattice, are proportional
to projection operators.  In a sense, the doubler is removed because
only one component can hop.   This choice for $r$ has been the most
popular in practice, but we remain more general.

The hopping parameter has a critical value at
$$
K_{crit}={M \over 2r} \neweq
$$
At this point the gap in the spectrum closes
and one species of Fermion becomes massless.  This is sketched in
Fig.~(4). The Wilson term, proportional
to $r$, still mixes the $a$ and $b$ type particles; so, there is no
exact chiral symmetry.  Nevertheless, in the continuum limit
this represents a candidate
for a chirally symmetric theory.  Beforehand, as
discussed in Ref.~[\refname\aokigockschref],
chiral symmetry does not provide a good order parameter.

\midinsert
\epsfxsize=.7\hsize
\noindent \narrower {{ \bf Fig.~(4)} The spectrum of Wilson Fermions
  at critical hopping.
  Here we take $K=0.5$, $M=0.25$, and $r=0.25$.}
\endinsert

A difficulty with this approach is that gauge
interactions will renormalize the parameters.  To obtain massless
pions one must finely tune $K$ to $K_{crit}$, an {\it a priori} unknown
function of the gauge coupling.  Despite the awkwardness of such
tuning, this is how numerical simulations with Wilson quarks generally proceed.
The hopping parameter is adjusted to get the pion mass right, and one
hopes for the remaining predictions of current algebra to reappear
in the continuum limit.

\bigskip
\noindent{\bf V. Supercritical $K$ and surface modes.}

The case of $K$ exceeding the critical value $M/2r$ is rarely
discussed but quite interesting nevertheless.  Aoki and Gocksch
[\aokigockschref] have argued that as one passes through this point
with gauge fields present, there
occurs a spontaneous breaking of parity, and if one has more than one flavor,
there is a breaking of flavor symmetry.  In their picture
one of the pion states becomes massless
from the critical behavior at $K_{crit}$, while the other two
remain as Goldstone bosons
of the broken flavor symmetry as $K$ increases still further.

\midinsert
\vskip -.4in
\epsfxsize=.6\hsize
\epsfysize=.5\vsize
\vskip -.5in
\noindent \narrower {{ \bf Fig.~(5)}
Energy levels versus hopping parameter $K$ for
Wilson Fermions on a 20
site lattice.  Here we take $M=1$ and $r=0.5$.  Note the low energy
states appearing when $K>K_{crit}=1$.}
\endinsert

Restricting ourselves to the free Fermion case for the time being,
interesting things happen here for supercritical $K$ as well.
As the band closes and reopens with increasing $K$, the
positive energy particle band and the negative energy Dirac sea
couple strongly.  A similar situation was
studied some time ago by Shockley [\refname\shockleyref], who
observed that if the system is finite with open walls, then two
discrete levels leave the bands and emerge bound to the ends of the system.
In Fig.~(5) we show this phenomenon by plotting
the spectrum of states for a box of 20 sites
as a function of the hopping parameter.

As the volume of our system goes to infinity,
particle-hole symmetry forces these surface levels
to go to exactly zero energy.  In a
finite box, the wave functions have exponential tails away from
the walls, mixing the states and in general giving them
a small energy.

In the appendix we prove the general result that
there exists such a state bound to any interface separating a
region with
$K>K_{crit}$ from a region with
$K<K_{crit}$.
In Ref.~[\kaplanref],
Kaplan uses $M=2Kr+m \epsilon(x)$.
We adopt the simpler approach of Shamir [\refname\shamirref]
and take $K=0$ on one
side, giving modes on an open surface.

The analysis of the appendix shows the essential nature of the side with
supercritical hopping.  In our later discussion of the
anomaly in terms of a flow into an extra dimension, it will always be a flow
into a region of supercritical hopping.  This should be contrasted
with the continuum discussion of Ref.~[\refname\callanharveyref], where the
flow is symmetric about the defect.

Following the usual procedure of filling half the states for the
Dirac sea, we see that there is an ambiguity with the last Fermion,
which could go into either of the degenerate surface modes.  If we
imagine coupling the Fermions to, say, a $U(1)$ gauge field, then
this last Fermion will be a source of a background electric field which
will run to the hole state on the opposite wall.  This is the physical
origin of the parity breaking proposed in Ref.~[\aokigockschref].
In the continuum limit the vacuum should be equivalent to that
of the massive Schwinger model with a half unit of background
electric flux.  The physics of this model in the continuum
was discussed in Ref.~[\refname\colemanbosonref].

\bigskip\noindent
{\bf VI. Extra dimensions}

As the system size goes to infinity, the surface modes quite
naturally go to zero energy.  This
behavior forms the basis for a theory of chiral Fermions.  The approach
of Kaplan [\kaplanref] is to reinterpret the
coordinate labeled by $j$ in the above discussion
as an extra dimension beyond the usual
ones of space and time.  Our physical world exists on
a four dimensional interface, with the light quarks and leptons
being the above surface modes.

To be concrete, let us consider adding $D$ space dimensions to the
above Hamiltonian, where for the following $D$ will either be 1 or 3.
For simplicity we will take $L^D$ space sites and use antiperiodic
boundary conditions for each of these dimensions.
In the extra dimension, which we refer to as the fifth, has $L_5$ sites
and open boundaries.
We take the same hopping and Wilson parameters in each of the
dimensions, including the fifth, although this is not essential.

Our Dirac matrices $\gamma_\mu$ satisfy the usual
$$
[\gamma_\mu,\gamma_\nu]_+=2g_{\mu\nu}.
\neweq
$$
We define $\gamma_5=i\gamma_0\gamma_1\gamma_2\gamma_3$ for $D=3$ and
$\gamma_5=\gamma_0\gamma_1$ for $D=1$.  We take $\gamma_0$ and
$\gamma_5$ to be Hermitian, while the spatial $\gamma$ matrices
are anti-Hermitian.
The Hamiltonian we are led to is then
$$
\eqalign{
H=\sum_{{\bf n},j}
\bigg(&K\overline\psi_{{\bf n},j+1}
(\gamma_5-r)\psi_{{\bf n},j}
-K\overline\psi_{{\bf n},j}
(\gamma_5+r)\psi_{{\bf n},j+1}\cr
+&\sum_{a=1}^D (K\overline\psi_{{\bf n}+{\bf e}_a,j}
(i\gamma_a-r)\psi_{{\bf n},j}
-K\overline\psi_{{\bf n},j}
(i\gamma_a+r)\psi_{{\bf n}+{\bf e}_a,j})\cr
+&M\overline\psi_{{\bf n},j}\psi_{{\bf n},j}\bigg).\cr
}\neweq
$$
Here $\bf n$ denotes the spatial sites, and ${\bf e}_a$ is the unit
vector in the positive $a$'th direction.

The use of antiperiodic boundary conditions allows us to go to momentum
space for the spatial coordinates.  Denoting the components of the momentum
by $q_a$, we write
$$
\psi_{{\bf q},j}={1\over
L^{D/2}}\sum_{\bf n}e^{-i{\bf q}\cdot{\bf n}} \psi_{{\bf n},j}.
\neweq
$$
Each component of the momentum takes discrete
values from the set $(2k+1)\pi/L$ where $k$ runs from, say,  0 to $L-1$.
This makes our Hamiltonian block diagonal, with each value
for ${\bf q}$ representing a separate block.  In this way the Hamiltonian
reduces to
$$
\eqalign{
H=\sum_{{\bf q},j}
\bigg(&K\overline\psi_{{\bf q},j+1}
(\gamma_5-r)\psi_{{\bf q},j}
-K\overline\psi_{{\bf q},j}
(\gamma_5+r)\psi_{{\bf q},j+1}\cr
+&\sum_a 2K\sin(q_a)\overline\psi_{{\bf q},j}\gamma_a\psi_{{\bf q},j}\cr
+&\big(M-2Kr\sum_a\cos(q_a)\big)\overline\psi_{{\bf q},j}\psi_{{\bf
q},j}\bigg).\cr
}\eqname{\bigHamiltonianeq}
$$

\midinsert
\vskip -.3in
\epsfxsize=.6\hsize
\epsfysize=.5\vsize
\vskip -.4in
\noindent \narrower {{\bf Fig. (6)} The energy spectrum as a function of
the physical momentum on a lattice with $L_5=11$ and $L=500$.
Note the crossing surface modes at low energy. Because of large spatial
size of the lattice, momentum appears to be almost a continuous variable.}
\endinsert

As follows from the discussion in Appendix, modes bound to the surface
in the fifth direction exist whenever $K$ exceeds the critical value
$$
K_{crit}=M/2r-K\sum_a \cos(q_a). \neweq
$$
Note how this critical value now depends on the spatial momentum.
Appropriately choosing $M$, we can have the surface states exist for small
$q$, but not when any component of $q_a\sim\pi$, thus avoiding
doublers [\refname\jansenschmaltzref].
Specifically, when the hopping is direction independent, we want
(assuming $K$, $r$, and $M$ are all positive)
$$
(D-1)K<M/2r<(D+1)K. \neweq
$$
For convenience we denote $\tilde M = M - 2KrD$ and, if not specified
otherwise, we use in the illustrative figures that follow
the typical set of parameters $\tilde M = 0.1, K = 0.6, r = 11/12$.
In Fig.~(6) we sketch the spectrum of the Hamiltonian (\bigHamiltonianeq)
in one space dimension, showing that we indeed avoid doublers in this
case.

The fact that the low energy states are bound to the surface
of the system is easily seen by studying the expectation value
for the fifth coordinate in the single particle states.
In Fig. (7) we show the energies of the various levels
as a function of this expectation on a $L=500$ by $L_5=11$ lattice.

\midinsert
\vskip -.35in
\epsfxsize=.6\hsize
\epsfysize=.5\vsize
\vskip -.4in
\noindent \narrower {{\bf Fig.~(7)} The energy spectrum as a function
of the average position in the extra dimension on
a $L_5=11$ by $L=500$ lattice.
Note how the low energy modes lie near the ends of the lattice.}
\endinsert

The above discussion shows that on a single surface we have an
elegant lattice theory for a low energy chiral Fermion.
We would now like to add gauge fields.  Here we adopt
the attitude that we do not want a lot of new degrees of
freedom, and follow Ref.~[\nnref] in
regarding the extra dimension as a flavor space.
In particular, we do not put gauge fields in the fifth dimension,
and the physical gauge fields are independent of this dimension.

While this approach has the advantage of preserving an exact
gauge invariance and not introducing lots of
unwanted fields, it has the disadvantage that both walls are coupled
equally to the gauge field.  Thus, even when the size of the fifth dimension
approaches infinity, the opposite chirality Fermions do not decouple.
The main thing that has been accomplished so far is to find a theory
of Fermions coupled in a vectorlike manner, without any doublers,
and with a natural way to take the Fermion masses to zero.

Later we will briefly discuss decoupling the states on one wall in a gauge
non-invariant way.  This may give rise to a theory of chirally coupled
Fermions, but the role of anomaly cancellation remains unclear.

\bigskip\noindent
{\bf VII. The anomaly and rotating eigenvalues}

	If one's goal is to formulate the massless Fermion theory
in $2n$ dimensions (chiral or vectorlike) using the surface
low energy states of the massive vectorlike theory in $2n+1$ dimensions,
then the question of anomalies is of primary concern. The reason
is quite obvious: If we look at one interface with one chiral
Fermion living on it, then we should see the anomaly in gauge current.
On the other hand, the full theory we started from is anomaly free.
Therefore, one should not only make sure that the correct anomaly
is reproduced, but also understand the mechanism of how this
happens.

	The basic scenario that clarifies the situation
was discussed in a somewhat different context by Callan and
Harvey [\callanharveyref]. They consider a vector theory,
whose mass term has a domain wall shape in an extra
dimension, and show that it has a chiral zeromode living on the
wall. The anomalous gauge current generated by this zeromode
has to be cancelled in the underlying $2n+1$ dimensional
theory since that world is anomaly free. Indeed, the massive modes
contribute to the low energy effective action a piece representing
the flow of charge into (or out of) the wall from the extra dimension.
When calculated far from the wall, it cancels the anomalous
contribution. In the $U(1)$ case this cancellation was recently
explicitly checked to be valid even close to the wall
[\refname\chandrasekharanref]. Therefore, what on the interface
looks like an anomaly is the flow of charge into the extra dimension
and the role of the heavy modes is to carry that charge.

The above picture was adopted by Kaplan in his lattice proposal
[\kaplanref] and in Ref.~[\refname\shamirprimeref] Shamir
carried out a detailed study of
the anomaly in this scheme.  Here we explicitly visualize
how this works with our Hamiltonian. We already mentioned
that at any finite $L$ the model is strictly vectorlike with our
gauge prescription.  Nevertheless, since opposite
chirality partners live on opposite walls, the charge still
has to be transported through the extra dimension and we should see
exactly what Callan and Harvey suggested.  We will
examine in detail how the heavy modes behave
in external fields, causing the anomaly.
As is often the case, the Hamiltonian formalism can offer
more  physical intuition than the (often practically more
powerful) Lagrangian approach. Since we want to see the heavy modes
``in action,'' we work
with the full Fermion theory and study how the one particle states respond
to a particular external field.

For simplicity, we concentrate on the one dimensional case
with gauge group $U(1)$.  If we have an external $U(1)$ gauge field,
it is manifested in terms of phase factors whenever
a Fermion hops from one site to the next.  Calling this factor
$U(k)=e^{i\alpha_k}$ for the hopping from space site $k$ to $k+1$, the
full Hamiltonian becomes
$$
\eqalign{
H=\sum_{k,j}
\bigg(&K\overline\psi_{k,j+1}
(\gamma_5-r)\psi_{k,j}
-K\overline\psi_{k,j}
(\gamma_5+r)\psi_{k,j+1}\cr
+&K\overline\psi_{k+1,j}U(k)
(i\gamma_1-r)\psi_{k,j}
-K\overline\psi_{k,j}U^\dagger(k)
(i\gamma_1+r)\psi_{k+1,j}\cr
+&M\overline\psi_{k,j}\psi_{k,j}\bigg)\cr
}\neweq
$$

A gauge transformation by a phase $g$ at site $k$ takes $U(k)$ to
$U(k)g^\dagger$, $U(k+1)$ to $g U(k+1)$, and $\psi_k$ to $g\psi_k$.
The invariance of the Hamiltonian under this symmetry tells us that
the spectrum only depends on the product of all the $U(k)$.  This is
the net phase acquired by a Fermion in travelling all the
way around our finite periodic system.

A particularly convenient gauge
choice is to evenly distribute the phases so that $U(k)=e^{i\alpha}$ is
independent of $k$.  This keeps momentum space simple,
with the Hamiltonian becoming
$$
\eqalign{
H=\sum_{q,j}
\bigg(&K\overline\psi_{q,j+1}
(\gamma_5-r)\psi{q,j}
-K\overline\psi_{q,j}
(\gamma_5+r)\psi_{q,j+1}\cr
+&2K\sin(q-\alpha)\overline\psi_{q,j}\gamma_1\psi_{q,j}\cr
+&(M-2Kr\cos(q-\alpha))\overline\psi_{q,j}\psi_{q,j}\bigg)\cr
}\neweq
$$
The energy eigenstates are functions of the momentum
shifted by $\alpha$.  As $\alpha$ increases by $2\pi/L$,
sequential momenta rotate into each other. The total net
phase in this case is $2\pi$ and, as expected, physics goes
back to itself. In what follows we will often refer to
$\alpha$ in its natural units, calling $2\pi/L$ one unit of flux.

In the adiabatic limit, the time evolution is a
continuous change of one particle states with changing $\alpha$.
It is therefore sufficient to consider the spectrum flow with
respect to $\alpha$ itself. In Fig.~(8) we show the positions of
one particle states (mean values of $x_5$) for the values
$\alpha=0,{1\over4},{1\over2},{3\over4}$. One can see quite transparently
what is happening; the low energy states at the lattice ends change energy
without
substantially changing
their position in the extra dimension.  The same is true for the very
high energy states, residing deep in the lattice interior.
However, the surface states with energies close to the cutoff
are very sensitive to the applied
field. When the energy of such a level gets close to the cutoff,
it moves swiftly towards the middle. At the same time, another
level from middle lowers its energy and runs towards the
opposite wall. This is also true for corresponding
levels with negative energy at the cutoff; they just move in the
opposite direction. Therefore, we see how the heavy modes right at the
cutoff are responsible for carrying the charge on and off the
end surfaces.  Consider applying the gauge field to the physical
vacuum with all negative energy states filled.  Then these
``flying states'' are responsible for what
appears to be the gauge anomaly on the surfaces.

\midinsert \vskip -.25in
\epsfxsize=.65\hsize
\vskip -.3in \noindent \narrower {{\bf Fig.~(8)}
The energy spectrum as a function of $<x_5>$ for
$\alpha=0,{1\over4},{1\over2},{3\over4}$ on a $L_5=11$ by $L=16$ lattice.
Successive positions of energy levels are marked by $1,2,3$ and $4$
respectively. The levels rotate in the anticlockwise sense.}
\endinsert

To make our discussion more formal and quantitative, we now compute
the anomaly in the axial charge of our vectorlike model.
To do that, we first give a simple definition of the axial charge.
On the lattice there is considerable freedom here;
any definition assigning opposite charges to the
two zeromodes living on the opposite walls should yield a correct
continuum limit. In our model, the most natural measure for ``being
chiral" is in fact the measure of ``being on the wall". Therefore,
we define the operator of the axial charge to be the Fermion number
weighted by the location in the extra dimension
$$
Q_5={1\over L_5-1}\sum_{q,j}(L_5-1-2j)
\psi^\dagger_{q,j}\psi_{q,j}, \neweq
$$
with $j=0, 1,2,...,L_5-1$.
This assigns to a one particle state $+1$ if it is exactly bound
to the left wall with $j=0$ and $-1$ if it is bound
to the right wall. For states smeared uniformly over $j$,
we obtain zero.  Regarding the extra dimension as an internal
space, we see that the axial charge is nothing but a particular
combination of flavor charges.

We define the vacuum of the theory in the usual way by filling
all the negative energy eigenstates.  We then measure the energy of
excited states relative to this state.  At zero field, the vacuum has
zero axial charge because of the mirror symmetry of the Hamiltonian
with respect to the middle of the extra space $(j\rightarrow L_5-1-j)$. To see
the anomaly, we evolve the vacuum in an adiabatic field, increasing the
value of $\alpha$ from $0$ to $1$ unit of flux and look for the change in the
total axial charge.

What we should see if the theory is to have the correct
anomaly structure can be deduced from the continuum expression
for chiral anomaly in two dimensions, namely
$$
\eqalign{
\Delta Q_5(t)& = {e \over 2\pi} \int_0^t dt^\prime \int dx
              \epsilon_{\mu\nu} F^{\mu\nu}(x,t^\prime)\cr
             & = -{e\over 2\pi} L \int_0^t 2{\dot A}_1(x,t^\prime)
               = -2e{L\over 2\pi}A_1(t).
}\neweq
$$

In our units and with identification $\alpha=eA$ we have
$$
\Delta Q_5(t)=-2\alpha (t)\neweq
$$
and what we should see is a straight line with the slope
of $-2$.

\midinsert
\vskip -.25in
\epsfxsize=.6\hsize
\epsfysize=.5\vsize
\vskip -.3in \noindent \narrower {{\bf Fig.~(9)}
The total axial charge of the vacuum as a function of $\alpha$. The connected
line represents the evolution for the time scale of turning on the
gauge field $\tau$ being much longer than the timescale $\delta$ set
by the particle tunnelling through the extra dimension.
The continuation by crosses applies for $0\ll\tau\ll1/\delta$.}
\endinsert

In Fig.(9) we show that this is indeed what happens in our model.
As the field is turned on, the levels in the Dirac sea start to move
anticlockwise $(\alpha>0)$, decreasing the total axial charge to
negative values in an almost strictly linear manner. In fact, a
least square fit yields the slope -1.995 in this
particular case with negligible admixture of higher powers.

When the value of the field is close to one half unit of flux, one of the
filled levels on the right surface is just about to become the positive
energy particle and one empty level on the left is just about to drop into
the sea. However, as long as our extra dimension is finite,
the surface states are not exactly massless.
There is always a tiny mass $\delta$
present, caused by mixing of the states on the opposite walls. When the
fields are truly adiabatic, with the typical time $\tau$ for turning on
the fields longer than any other time scale in the problem, e.g.
$\tau\gg 1/\delta$, instead of creation of a particle-hole pair, the
two levels will have enough time to exchange between the walls.
As a consequence,
we observe a jump at $\alpha=1/2$, which allows vacuum to evolve into
its original state as $\alpha$ approaches one unit of flux.  In Fig.(8)
this is represented by the two almost degenerate levels with energies
$\pm\delta$, located in the middle of the pattern.
Although the
total change of the axial charge is zero in this case, the anomaly can
still be extracted from the slope for fields smaller than one half unit.
Note also that for all practical purposes, the time scale $1/\delta$
is so huge (in our case we have $\delta \approx 10^{-9}$ with just
$11$ sites in the extra dimension) that only an extremely slow turning on
of the field can be rightfully considered adiabatic.
With a faster turn on, a
particle-hole pair is created on opposite walls, and at one unit of flux
the net change in $Q_5$ approaches $-2$, as it should.
In Fig.~(9) this behavior is represented by pluses continuing
on the straight line.

\bigskip\noindent{\bf VIII. Chiral models with broken gauge invariance}

In an attempt to construct a truly chiral model we would like to
decouple the zeromodes located on the opposite walls. The common
gauge field shared by the two walls couples the chiral partners directly
to each other.  To avoid this, we
would like the gauge fields to be felt by one of the zeromodes
but not by the other; we would like to switch off
the gauge field as one traverses the extra dimension.

The problem is that when the gauge field on a given link at the $j$-th
slice in the extra dimension is different from the value
at the slice $(j+1)$, local gauge invariance is broken at that
particular place. Any ``switching off'' procedure is therefore
unavoidably accompanied by an explicit gauge symmetry breaking in the
full theory on a finite lattice.

There are many ways to turn off the field. One natural choice
is singled out by the requirement that the gauge symmetry is broken
in the smallest region possible, namely only between two slices
in the extra dimension. In this way we are led to study the model,
consisting of $N=N_1+N_2$ slices in extra dimension, with $N_1$ slices
coupled to the same gauge field as before and with zero field on the
remaining $N_2$ slices. In what follows we consider even number
of sites in the extra dimension and symmetric case $N_1=N_2=N/2$.

This scheme is closely related to that used in [\refname\gjpvref].
In that paper an additional scalar field was introduced to restore
a local gauge invariance.  This was analyzed in the small field
regime, where it was found that new low mass modes were bound to the
location of the gauge field shutoff in the extra dimension.  Here
we consider effectively freezing this scalar field to a unit value.

\midinsert
\vskip -.25in
\epsfxsize=.6\hsize
\epsfysize=.5\vsize
\vskip -.3in
\noindent \narrower {{\bf Fig.~(10)}
Energy spectrum as a function of $x_5$ for zero field (black points)
and for $\alpha=-1/2$ (pluses), sharply switched off at the middle
of the extra-space. In this case we consider $L_5=12$ by $L=24$,
lattice. The low energy states on the wall with nonzero
field move up while those on the opposite wall stay unchanged.}
\endinsert

Another possibility would be to gradually turn off the gauge field
as we pass through the extra dimension.  This corresponds to letting
the gauge field leak away as one goes deeper into the slab.  Our results
with such a scheme were qualitatively similar to what we show below for
a sharp cutoff.  On the other hand, it is possible that ambiguities
might arise with a gradual cutoff of non-abelian fields.

It is unclear what in principle is wrong with sacrificing
gauge invariance at this stage. Indeed, we would be happy
if the net result were a well regulated theory of single chiral
Fermion. The theory seems to behave as desired when classical
external fields are considered, as we will see implicitly below.
Nevertheless, there are plausibility arguments that, when the gauge
fields are dynamical, gauge invariance may be dynamically restored
with the theory becoming vectorlike again (see [\shamirref] and
references therein).

\midinsert
\vskip -.25in
\epsfxsize=.6\hsize
\epsfysize=.5\vsize
\vskip -.3in
\noindent \narrower {{\bf Fig.~(11)}
Energy spectrum as a function of momentum for the field equal to minus
one unit of flux, sharply switched off at the middle of the extra-space.
We show the results on a $L5=12$ by $L=24$ lattice, indicating that
one level (right-mover) escaped the Dirac sea.}
\endinsert

Here we do not fully resolve this question. Numerical
simulations may be necessary to see in detail what is
going on. Instead, we point out one amusing possibility that
this gauge variant model seems to offer. Namely, that it gives
rise to a natural definition of a winding number for
classical gauge configurations on a lattice. This involves counting
the quantum one-Fermion states of the gauge variant system
in a given external gauge field.

The basic idea is as follows.
Assume that the field resides on the left half of the slab, while on
the right half the field is permanently turned off. For a given
external field, we can diagonalize the Fermion Hamiltonian
to obtain a set of one particle levels.
If we start from zero field and increase the field gradually, one expects
the levels located on the left to rotate in a similar way to that
seen in the previous section for
the gauge invariant model. Eventually, one positive level will
cross zero and become a negative energy level. On the other hand,
the levels on the right will not feel these changes and are not
expected to rotate. The crossing level on the left is not compensated by a
crossing partner (in the opposite direction) on the right and as a net
result we have one more level in the Dirac sea. Increasing the field
further, more levels will go down, while switching the field to
negative values should do the opposite: expel the states out of the
sea. In this way, we divide the gauge configurations into
classes, each class labeled by the ``excess of levels in the
Dirac sea.''

In Figures (10) and (11) we demonstrate that the gauge variant model
indeed behaves as described above. Fig.~(10) shows the shift
of levels as the field is turned on from zero value to minus half unit.
The low energy levels, located on the left move up and one of them
is just about to escape the Dirac sea. The levels on the opposite
wall on the other hand stay unchanged and this behavior is to be
compared with the one in gauge invariant model (see Fig.~(8)).
Fig.~(11) shows clearly how the vacuum is deformed when minus one unit
of flux is turned on, expelling one right-moving level.

\midinsert
\vskip -.25in
\epsfxsize=.6\hsize
\epsfysize=.5\vsize
\vskip -.3in
\noindent \narrower {{\bf Fig.~(12)}
Winding number as defined in the text for several gauge field
configurations (values of $\alpha$). The staircase assignment takes
place as expected.}
\endinsert

Formally, we can define the winding number by
$$
n={1\over2}(N_- - N_+),\neweq
$$
where $N_+,N_-$ is the number of positive and negative levels
respectively. In Fig.~(12) we show that with this definition
the expected staircase assignment of the winding number really
takes place.

Note that while their behavior is rather complex,
none of the levels deep in the middle of the extra dimension
drop to low energy with these small values for the field.  With our strong
breaking of gauge invariance, we avoid the low energy states seen in
Ref.~[\gjpvref] bound to the place where the gauge field is cut off.

Except for exceptional cases where an eigenvalue
exactly vanishes, this definition is always a well defined integer.
Note that this is true regardless of whether the gauge field
represents a gauge copy of the vacuum or not.
An instanton in this language would occur whenever a gauge field
configuration interpolates in time between two spatial vacuum
configurations of different winding number [\cmtref].

The correspondence with the continuum case is immediate.
Consider continuum massless electrodynamics on a circle with
circumference $L$ and pick the gauge $A_0(x,t)=0, A_1(x,t)=A_1(t)$.
The periodic gauge transformations with winding number $n$ are
specified by $\Lambda=e^{inx2\pi/L}$ and the pure gauge configurations
generated by these functions
are $A_1=n2\pi/L=-i\Lambda^{-1} {d \over dx} \Lambda$. This is
what we see for integer $\alpha$ in Fig.(12).  It might be interesting
to compare this definition of winding number with other
prescriptions, i.e. Ref.~[\refname\cghnref].

\bigskip\noindent
{\bf IX. Weak interactions, mirror Fermion model}

With an exact gauge invariance and a finite size for the extra dimension,
the surface models are inherently vectorlike.  The Fermions always appear
with both chiralities, albeit separated in the extra dimension.  However,
experimentally we know that only left handed neutrinos couple to
the weak bosons.  In this section we discuss one way to break the symmetries
between these states, resulting in a theory with only one light
gauged chiral state.  Here we keep the underlying
gauge symmetry exact, but do
require that the chiral gauge symmetry be spontaneously
broken, just as observed in the standard model.  The picture also contains
heavy mirror Fermions.  If anomalies are not cancelled amongst the light
species, these heavy states must survive in the continuum limit.  It remains
an open question when anomalies are properly cancelled whether
it might be possible to drive the
heavy mirror states to arbitrarily large mass.

We start by considering
two separate species $\psi_1$ and $\psi_2$ in
the surface mode picture.  However, we treat these in an unsymmetric
way.  For $\psi_1$ use our previous Hamiltonian.  For
$\psi_2$ we change the sign of all terms proportional to $\gamma_5$.
On a given wall, the surface modes associated with $\psi_1$ and
$\psi_2$ will then have opposite chirality.

Now we introduce the gauge fields.
Since we want to eventually couple only one-handed neutrinos to the
vector bosons, consider gauging
$\psi_1$ but not $\psi_2$.  Indeed, at this stage $\psi_2$ represents
a totally decoupled right handed Fermion on one wall.  We still have a
mirror situation on the opposite wall, consisting of a right
handed gauged state and a left handed decoupled Fermion.

The next ingredient is to spontaneously break
the gauge symmetry, as in the standard
model, by introducing a Higgs field $\phi$ with a non-vanishing
expectation value.  We can use this field to
generate masses as in the standard model by coupling $\psi_1$ and
$\psi_2$ with a term of the form $\bar\psi_1\psi_2\phi$.

The new feature is to allow
the coupling to the Higgs field to depend on the extra coordinate.
In particular, let it be
small or vanishing on one wall and large on the other.
The surface modes are then light on one wall and heavy on the other.

In Fig.~(13) we consider one space dimension and
sketch the Fermion spectrum of this model with vanishing gauge
fields and a constant Higgs field.  With an increasing gauge field as
considered in the previous section, only one component of the light
Fermions will shift.  As the flux approaches one half unit, however, something
new happens.  Now the Higgs field will also be affected,
and it can become favorable for it to acquire a winding to reduce
the system energy.  As the flux approaches a full unit, then the
physics of the model returns to a gauge transform of the original vacuum.

This model is quite closely related to the picture of the previous section.
There we could have restored the gauge invariance with a Higgs field, as
was done in Ref.~[\gjpvref].  Then the model is essentially equivalent to
that considered here.  The difference in the last section was
that we froze the Higgs field and did
not allow the above unwinding to occur.  This broke the gauge symmetry,
and the physics with a full unit of flux is not equivalent to the
gauge invariant situation discussed here.  On the other hand, the
explicit breaking was essential to the definition of winding number.

\midinsert
 \vskip -.4in
\epsfxsize=.6\hsize
\epsfysize=0.5\vsize
\vskip -.5in
\noindent \narrower {{ \bf Fig.~(13)}
Energy levels versus momentum for
the two species model discussed in the text. In this case we take
$L_5=10$ by $L=60$ lattice and switch off the coupling between species
sharply at the middle of the extra space. Note how one species is
massless and the other massive.}
\endinsert

As in other mirror Fermion models [\refname\montvayref], triviality
arguments suggest that there might exist bounds on the mass of the
heavy particles.  This is certainly expected to be the case where
the light Fermions alone give an anomalous gauge theory, in which
case we expect the mirror particles cannot become much
heavier than the vector mesons, i.e. the $W$.

It is conceivable that the restrictions on the mirror Fermion masses
are weaker when anomalies cancel amongst the light states.  In this case
there is no perturbative need for the heavy states, and
perhaps they can be driven to infinite mass in the continuum limit.
If so, we would have a candidate for a lattice discretization of the
standard model.

Unfortunately, the model as it stands does not lead to baryon
number violation [\refname\distlerreyref].  The anomaly will involve a
tunnelling of baryons
from one wall to the opposite, where they become mirror baryons.  Even
if these extra particles are heavy, the decay can only occur through
mixing with the ordinary particle states.   In this sense, the
mirror particles still show their presence
in low energy physics.

A speculative proposal is to use the right handed mirror states in
some way as observed particles.  Indeed, the world has
left handed leptons and right handed antibaryons.
Any simple extension of this idea to a realistic model must
unify these particles [\refname\unifiedref].  On the other hand,
the fact that the anomalies are canceled between different
representations of the $SU(3)$ of strong interactions may
preclude such options.

\bigskip\noindent{\bf X. Conclusions}

We have studied the use of Shockley surface states as the basis of
a theory of chiral Fermions.  For strong interaction physics this
yields an elegant formulation where the massless limit for the quarks
is quite natural.

We have seen explicitly how the anomaly works in terms of a flow in
the extra dimension used to formulate the model.  For anomaly free currents,
the net flow in this dimension cancels, and we expect the predictions of
current algebra to arise naturally.  On the other hand, the symmetries
for singlet axial currents are strongly broken by this flow.
This presumably precludes the
need for a corresponding Goldstone boson and solves
the $U(1)$ problem.

Several questions remain before we have a theory
of the weak interactions on the lattice, where the gauge fields
are to be coupled to chiral currents.  We seem to be led to a theory with
mirror Fermions on the opposing walls of the system.  In a spontaneously
broken theory these extra states can be given different masses.  Whether
they can be driven to infinite mass in the continuum limit presumably
depends on whether all necessary chiral anomalies have been cancelled.

\bigskip\noindent
{\bf Appendix: Conditions for zero modes}

In this appendix we prove the general result that a Fermion zero mode
will be bound to any defect separating a region with supercritical
hopping from a region with subcritical hopping.
We restrict ourselves to a single spatial momentum ${\bf q}$
from the full Hamiltonian of Eq.~(\bigHamiltonianeq), and look
for single particle states of the form
$$
\vert\chi\rangle=\sum_j \psi^\dagger_j\chi_j\vert 0\rangle
\neweq
$$
Here $\chi$ represents a spinor of ordinary numbers (i.e. it does
not anticommute with anything), and the vacuum $\vert 0\rangle$
is the state annihilated by the operator $\psi$.

We are interested in eigenstates satisfying the Schroedinger
equation
$$
H\vert\psi\rangle=E\vert\psi\rangle. \neweq
$$
We later consider position dependent values for the parameters
$K$, $M$, and $r$, but for now consider working in a region
where they are constant.
Since our Hamiltonian is then translation invariant, we first study
solutions of the form
$$
\chi_j\sim \lambda^j \chi_0 \neweq
$$
with $\lambda$ being a general complex number.
For plane waves we would restrict ourselves to
$\lambda=e^{iq}$; however, as we later plan
to match exponential solutions across defects, we remain more general.

Putting this all together, we find that the eigenvalues and the
wave function satisfy
$$
\left(\slash p + K(\lambda-{1\over\lambda})\gamma_5
+rK(\lambda+{1\over\lambda}) - (M-2Kr\sum_a\cos(q_a))\right )\chi_0=0
\eqname\eigeneq
$$
where we have defined $p_0=E$, $p_a=2K\sin(q_a)$, and the usual
$\slash p=p_0\gamma_0-\sum_a p_a \gamma_a$.

Multiplying Eq.~(\eigeneq) by
$\slash p + K(\lambda-{1\over\lambda})\gamma_5
- rK(\lambda+{1\over\lambda}) + (M-2Kr\sum_a\cos(q_a))$
tells us that any non-trivial solution must satisfy
$$
p^2+K^2\left(z^2-4-(M/K-rz-2r\sum_a\cos(q_a))^2\right)=0 \neweq
$$
where we define $z=\lambda+1/\lambda$.
Note that for any solution $\lambda$, $1/\lambda$ also is.
Thus for every exponentially increasing solution, there is
another which exponentially decreases.

We are ultimately interested in chiral solutions representing massless
particles. This suggests we look for states satisfying
$$
\slash p \chi=0\neweq
$$
For vanishing spatial momentum this is equivalent to $E=0$.
This restriction simplifies the equations dramatically.
Indeed, Eq.~(\eigeneq) then
implies that $\chi_0$ is then an eigenvector of $\gamma_5$.  Since this
matrix only has eigenvalues $\pm 1$, we conclude that
$$
\pm K(\lambda-1/\lambda) - M
    + Kr(\lambda+1/\lambda + 2\sum_a\cos(q_a)) = 0, \neweq
$$
which is a simple quadratic equation for lambda.
For a particular set of parameters we show the solutions
for $\lambda$ as a function of the hopping parameter $K$ in Fig.~(A1).
Some qualitative details of this picture apply
to small $r$, but a similar discussion applies in other cases.

\midinsert
\epsfxsize=.7\hsize
\noindent \narrower {{ \bf Fig.~(A1)} Translation eigenvalues at zero
energy versus hopping parameter. Note the crossing at the critical
hopping $K=1$. Here we have taken $r=.25$ and $M-2Kr\sum_a\cos(q_a)=.5$.
The plusses correspond to $\gamma_5\chi=\chi$ and
the boxes to $\gamma_5\chi=-\chi$.}
\endinsert

Note the crossing with $\lambda=1$ in Fig.~(A1) at the critical hopping.
This is a crucial point to the following. When $K$ exceeds the critical
value, both solutions  which satisfy $\gamma_5\chi=\chi$
have $\vert\lambda\vert<1$. Conversely, the two increasing wave
functions have the opposite chirality. This
means that when $K>K_{crit}$, if we can insure that a wave function
satisfies the chiral relation $\gamma_5\chi=\chi$,
it must exponentially decrease as $j$ increases.
We now discuss how to guarantee that this chiral condition is preserved.

Consider a defect region where the parameters vary with
position. Indexing  $M$ by its respective site and $K$ and $r$ by
the site on their left, the Schroedinger equation reads
$$
K_j(\gamma_5+r_j)\chi_{j+1}-K_{j-1}(\gamma_5-r_{j-1})\chi_{j-1}+
\left(\slash p_j -
M_j + 2K_jr_j\sum_a\cos(q_a)\right )\chi_j=0
\eqname\propagateeq
$$
Given $\chi$ on two adjacent sites, this determines its value on the
next site.  Thus we can iteratively propagate the wave function
from a pair of neighboring sites to any other location.
Suppose we start somewhere with a wave function which
satisfies $\slash p_j\chi_j=0$ and $\gamma_5 \chi_j=\pm  \chi_j$.
The important point is that Eq.~(\propagateeq) implies
that both these properties are propagated as we
move through the lattice.

To be explicit, consider the case where at large negative $j$
the hopping is subcritical, for large positive $j$ it is
supercritical, and there is some arbitrary transition region between.
If for large negative $j$ we start off with
the exponentially increasing solution with $\slash p\chi_j=0$  and
$\gamma_5 \chi_j=\chi_j$, then this solution can only couple
to the eigenvalues with the same chiral property at large positive $j$.
As discussed above, these both are exponentially decreasing.
Such a solution will be automatically normalizable.

The above argument shows that at least one state must exist bound to a region
separating sub from super critical hopping.
There could in principle exist more,
for example if the intervening region contains several domain walls.

\noindent{\bf References}

\item{\currentalgebraref.} S. Treiman, R. Jackiw, B. Zumino, and E. Witten,
{\sl Current algebra and anomalies,} (World Scientific, 1985, QC793.3.A4C87).

\item{\petcherref.} D.~Petcher, Nucl.~Phys.~B (Proc.~Suppl.) 30 (1993) 50.

\item {\thooftref.} G.~'t Hooft, Phys.~Rev.~Lett.~37, 8 (1976);
Phys.~Rev.~D14 (1976) 3432.

\item{\kaplanref.}  D.~Kaplan, Phys.~Lett.~B288 (1992) 342; M.~Golterman,
K.~Jansen, D.~Kaplan, Phys.~Lett.~B301 (1993) 219.

\item{\jansenref.} K.~Jansen, Phys.~Lett.~B288 (1992) 348.

\item {\latref.} M.~Creutz and I.~Horvath, Proceedings of Lattice 93
(in press).

\item{\cmtref.} M.~Creutz, I.~Muzinich, and T.~Tudron, Phys.~Rev.~D17
(1979) 531.

\item{\ambjornref.} J.~Ambjorn, J.~Greensite, and C.~Peterson,
Nucl.~Phys.~B221 (1983) 381; B.~Holstein, Am.~J.~Phys.~61 (1993) 142.

\item{\nogoref.} H.~Nielsen and M.~Ninomiya, Nucl.~Phys.~B185 (1981) 20;
B193 (1981) 173.

\item{\nnref.} R.~Narayanan and H.~Neuberger, Phys.~Lett.~B302 (1993) 62;
preprint RU-93-34; preprint RU-93-25.

\item{\wilsonref} K.~Wilson, in {\sl New Phenomena in Subnuclear Physics},
edited by A.~Zichichi (Plenum Press, N.~Y., 1977).

\item {\aokigockschref.} S.~Aoki, Nucl.~Phys.~B314 (1989) 79;
S.~Aoki and A.~Gocksch, Phys.~Rev.~D45 (1992) 3845.

\item{\shockleyref.} W.~Shockley, Phys.~Rev.~56 (1939) 317; see also
F.~Seitz,
{\sl The Modern Theory of Solids}, (McGraw-Hill, 1940), p.~323-4;
W.~G.~Pollard, Phys.~Rev.~56 (1939) 324.

\item {\shamirref.} Y.~Shamir, Nucl.~Phys.~B406 (1993) 90.

\item {\callanharveyref.} C.~Callan and J.~Harvey, Nucl.~Phys.~B250
 (1985) 427.

\item {\colemanbosonref.} S.~Coleman, Annals Phys. 101 (1976) 239.

\item {\jansenschmaltzref.} K.~Jansen and M.~Schmaltz, Phys.~Lett.~B296 (1992)
374.

\item {\chandrasekharanref.} S.~Chandrasekharan, Columbia U.~preprint
CU-TP-615 (1993).

\item {\shamirprimeref.} Y. Shamir, Weizmann preprint WIS-93-99-PH (1993).

\item {\gjpvref.} M.~Golterman, K.~Jansen, D.~Petcher, and J.~Vink,
preprint UCSD-PTH-93-28 (1993).

\item {\cghnref.} M. Chu, J. Grandy, S. Huang, and J. Negele, preprint
MIT-CTP-2269 (1993).

\item {\montvayref.} I.~Montvay, Nucl.~Phys.~B (Proc.~Suppl.) 30 (1993) 621;
Phys.~Lett.~199B (1987) 89.

\item {\distlerreyref.} J. Distler and S. Rey, Princeton
preprint PUPT-1386 (1993).

\item {\unifiedref.} S.~Frolov and A.~Slavnov, Max-Planck
Inst.~preprint MPI-Ph 93-12 (1993);
S.~Aoki and Y.~Kikukawa, preprint UTHEP-258 (1993).

\vfill
\eject
\bye